\documentclass[twocolumn,showpacs,preprintnumbers,amsmath,amssymb]{revtex4}
\usepackage{graphicx}
\usepackage{dcolumn}
\usepackage{bm}

\begin{document}

\title{Stacking Interactions in Denaturation of DNA Fragments}

\author{ Marco Zoli }
\affiliation{
School of Science and Technology - CNISM \\  Universit\`{a} di Camerino, I-62032 Camerino, Italy \\ marco.zoli@unicam.it}

\date{\today}

\begin{abstract}
A mesoscopic model for heterogeneous DNA denaturation is developed in the framework of the path integral formalism. The base pair stretchings are treated as one-dimensional, time dependent paths contributing to the partition function. The size of the paths ensemble, which measures the degree of cooperativity of the system, is computed versus temperature consistently with the model potential physical requirements. It is shown that the ensemble size strongly varies with the molecule backbone stiffness providing a quantitative relation between stacking and features of the melting transition. The latter is an overall smooth crossover which begins from the \emph{adenine-thymine} rich portions of the fragment. The harmonic stacking coupling shifts, along the $T$-axis, the occurrence of the multistep denaturation but it does not change the character of the crossover. The methods to compute the fractions of open base pairs versus temperature are discussed: by averaging the base pair displacements over the path ensemble we find that such fractions signal the multisteps of the transition in good agreement with the indications provided by the specific heat plots.
\end{abstract}


\maketitle

\section*{1. Introduction}

It has been long recognized that the physical properties of the DNA molecule are key to understand its biological function \cite{schro}. The core of the double helix embodies the genetic code through the sequence of base pairs. Gene transcription occurs as the hydrogen bonds between complementary bases can be broken thus leading to the formation of a transcription bubble in which the bases are exposed for chemical reactions \cite{wart}. While in heterogeneous fragments the bubble size is larger when the content of weakly bonded \emph{adenine-thymine} (AT) base pairs is higher, bubble formation can be also achieved experimentally by gradually heating the system as far as the two strands eventually separate. This process, known as thermal denaturation or melting, has been extensively studied in the last decades due to its relevance for the comprehension of the replication and transcription mechanisms \cite{yakus}. However, agreement has not been reached so far regarding the character of the melting transition whether first- or second- order. Besides being intriguing for the statistical physicists community the question has biological importance also in view of recent investigations pointing to correlations between thermodynamical melting properties and coding sequences in genomes \cite{yera,carlon1,jost}. Thermal denaturation is exploited in molecular biology \cite{gilli} where an understanding of the sequence specificity of melting is desirable for polymerase chain reaction.

Two families of  theoretical approaches have been developed to account for the melting phenomenon. The \emph{first} is based on the Poland-Scheraga model \cite{poland,azbel1,fisher,richard} which sees the DNA as a two state Ising-like sequence of base pairs with regions of variable size, denaturated loops, opening temporarily due to thermal fluctuations. In this picture self-avoiding interactions between loops and the rest of the helix \cite{carlon} may sharpen the denaturation leading to a first order transition in two dimensions and above \cite{peliti,stella} while a mechanically induced denaturation should remain second order \cite{hanke}. Improvements of Ising-like models, introducing rotational degrees of freedom and accounting for bending rigidity of bubbles, reduce the importance of loop entropy and also suggest that the denaturation is rather a smooth crossover \cite{manghi1}.  The \emph{second} approach assumes the Peyrard-Bishop (PB) model \cite{pey1} in which the stretching between complementary base pairs is represented by a one dimensional, continuous variable bringing the advantage that also intermediate states in the DNA dynamics can be described.  As a key refinement of the PB Hamiltonian, anharmonic stiffness has been introduced in the intra-strand stacking potential \cite{pey2} thus accounting for an entropy driven sharp denaturation \cite{theo}. While the PB model has proved successful to predict multistep melting in heterogeneous DNA \cite{cule,pey3}, several methodologies such as molecular dynamics, transfer integral calculations \cite{joy09} and renormalization group techniques \cite{santos1} have been employed to investigate the nature of the transition but the question is still unsettled.

DNA theories have to account for the fact that, given a base pair, the proton forming hydrogen bonds may be exchanged  between the pair mates whereas proton exchange may not occur in the adjacent base pair.  This amounts to say that base pair opening, the breathing of DNA \cite{gueron}, is a highly localized phenomenon. Accordingly the stacking coupling should be weak enough to allow for large conformational changes and high flexibility along the molecule backbone \cite{yakov,spassky}.
Moreover, even when no enzymes or thermo-mechanical forces are at work \cite{barb,seno}, the molecule undergoes large amplitude displacements showing the nonlinear nature of the double helix \cite{heeger,proh,pey5}. Nonlinearity is intrinsic to DNA and should not be attributed to heterogeneity as it persists also in artificial homogeneous molecules. Again, fluorescence methods have proved that large fluctuational openings take place in DNA hairpins while, for double stranded DNA, breathing fluctuations induce bubbles of a few base pairs which open spontaneously and can be monitored in real time \cite{bonnet1}.
Thus dynamical models for bubble formation and denaturation should incorporate fluctuation effects \cite{metz06,metz07}.

In this regard, a computational method based on the imaginary time path integrals \cite{fehi} has been recently proposed
\cite{io} to investigate the DNA thermodynamics in the temperature range for which denaturation is expected to occur. The base pair stretchings with respect to the ground state are thought of as time dependent paths with the time being an inverse temperature in the spirit of the Matsubara's Green functions formalism \cite{kleinert}.  The partition function for a nonlinear PB Hamiltonian is computed by summing, in the path configuration space, over an ensemble of path fluctuations around the ground state consistent with the model potential physical constraints. The denaturation appears as a rather smooth crossover both in homogeneous and heterogeneous DNA fragments \cite{io1} and, in the latter, the AT- rich regions drive a multistep transition as a function of temperature.
Precisely for the same sequences considered in Ref.\cite{io1}, a recent analysis \cite{singh} based on the extended transfer matrix approach \cite{zhang} finds consistent results regarding the temperature location of the main peaks in the specific heat profiles.

Although AT-rich bubbles form at lower temperatures, double helix openings extend also well inside the \emph{guanine-cytosine } (GC) domains indicating a role for nonlocal effects in shaping multistep denaturation patterns \cite{rapti,santos}. The sequence pattern is particularly relevant in segments made of a few tens of base pairs. This short scale, key to those transcription starting domains where the genes are read, is considered in this study.

While the path integral method has proved effective in dealing with a large number of base pairs displacements, modeling the denaturation still presents several unsolved questions even for a single molecule of finite length \cite{pey6}.
Provided that a molecule state corresponds to a point in the configuration space, statistical averages of physical quantities should be obtained in principle by summing over \emph{all} states with their Boltzmann weight. In practice numerical methods always require a confinement of the configuration space  \cite{joy05,ares} and, in the path integral method, such restriction naturally occurs by selecting the ensemble of possible paths for the base pair elongations.
The size of the ensemble depends however on the model potential.

In this paper, I develop the analysis carried out in Ref. \cite{io1} and investigate the effects of the backbone stacking interactions \emph{both }in shaping the path configuration space for a DNA sequence \emph{and} in determining the fractions of open base pairs versus temperature.  The Hamiltonian model is reviewed in Section 2 and the path integral method is described in Section 3. The results for a short heterogeneous fragment are presented in Section 4 where, in the denaturation temperature range, I calculate: \emph{a}) the fractions of open base pairs, \emph{b)} the free energy derivatives and \emph{c)} the size of the path ensemble by varying the strength of the backbone stacking. The confinement of the path amplitudes is also discussed by varying the cutoff in the path integration. Some Conclusions are drawn in Section 5.

\section*{2. Nonlinear Hamiltonian Model}

The nonlinear PB Hamiltonian \cite{pey2} for a chain of N heterogeneous base pairs reads

\begin{eqnarray}
& & H =\, \sum_{n=1}^N \biggl[ {{\mu \dot{y}_{n}^2} \over {2}} +  V_S(y_n, y_{n-1}) + V_M(y_n) \biggr] \, \nonumber
\\
& & V_S(y_n, y_{n-1})=\, {K \over 2} g(y_n, y_{n-1})(y_n - y_{n-1})^2 \, \nonumber
\\
& & g(y_n, y_{n-1})=\,1 + \rho \exp\bigl[-\alpha(y_n + y_{n-1})\bigr]\, \nonumber
\\
& & V_M(y_n) =\, D_n \bigl(\exp(-a_n y_n) - 1 \bigr)^2 \, ,
\label{eq:1}
\end{eqnarray}

where $y_n$, the transverse stretching at the {\it n-th} site, is the degree of freedom for the one dimensional model \cite{wart} and measures the relative pair mates separation from the ground state position. Although Eq.~(\ref{eq:1}) does not describe the actual geometry of the molecule, it contains the main contributions which energetically compete and determine the melting transition.
$\mu$ is the reduced mass of the bases which is taken identical both for GC- and AT- base pairs \cite{blake}. Consistently the backbone harmonic coupling $K$ is site independent: $K=\, \mu \nu^2$ with $\nu$ being the harmonic phonon frequency. Also the non linear (positive) parameters $\rho$ and $\alpha$ in the stacking potential $V_S(y_n, y_{n-1})$ are independent of the type of bases at $n$ and $n-1$. While the base pair stacking interactions generally stabilize the double helix \cite{yakov,spassky}, I have checked that the homogeneity assumption along the molecule backbone has scarce effect on the thermodynamics in the denaturation range. Instead, the latter is essentially determined by the size of $K$ as it will be discussed below.
Although the form for $V_S(y_n, y_{n-1})$ in Eq.~(\ref{eq:1}) may not be unique and other potentials have been proposed in the literature \cite{cule,joy05}, anharmonic stacking described by $\rho$ and $\alpha$ is key to the model as it establishes the link with the cooperative character of the bubble formation. Widely used values are taken in the following calculations, $\rho=\,1$  and $\alpha=\,0.35 {\AA}^{-1}$  \cite{theo,joy08}.

In fact, when the molecule is closed, $y_n \,, y_{n-1} \ll \alpha^{-1}$ for all $n$, the effective stacking coupling is $K(1 + \rho)$ and a large contribution to the stacking interaction
comes from the overlap of the $\pi$ electrons of the base plateaus. Whenever either $y_n > \alpha^{-1}$ or $y_{n-1} > \alpha^{-1}$, the corresponding hydrogen bond breaks and the base moves out of the stack thus reducing the electronic overlap. Accordingly  the interaction between neighboring bases along the strand weakens and also the next base tends to open. Formally, in Eq.~(\ref{eq:1}), $g(y_n, y_{n-1}) \sim 1$ and the effective stacking coupling between neighboring bases drops to $K$. This is the microscopic picture for the interplay between anharmonicity and cooperativity which determines the formation of a region with open base pairs. When the stacking gets weaker, the two strands become more flexible and their entropy increases. The actual rate of such increase approaching denaturation and above will determine the character of the transition.

Hydrogen bonds linking the two bases on complementary strands are modeled by the Morse potential $V_M(y_n)$ in Eq.~(\ref{eq:1}) \cite{proh1} which also incorporates the effects of heterogeneous sequences through $D_n$ and $a_n$.
$D_n$ is the dissociation energy of the pair: for DNA, energies per hydrogen bond are usually taken in the range $\sim 15 - 25 meV$. As AT- and GC- base pairs have two and three bonds respectively, I take $D_{AT}=\,30 meV$ and $D_{GC}=\,45 meV$. $a_n$ is an inverse length defining the spatial scale of the potential: transverse stretchings  for GC  are stiffer than for AT base pairs \cite{campa}. Accordingly I set $a_{AT}=\,4.2 {\AA}^{-1}$ and $a_{GC}=\,5 {\AA}^{-1}$ while slightly different values can be found in the literature.

$V_M(y_n)$ is standard for chemical bonds and reproduces the main properties of the inter-strand forces: \emph{a)} it has a hard core ($y_n < 0$) corresponding to the repulsion of the charged phosphate groups of the backbone; \emph{b)} it has a minimum ($y_n = 0$) for the ground state equilibrium distance between the two bases; \emph{c)} it has a plateau for large $y_n$ accounting for the fact that the force between the bases vanishes at the dissociation energy. Certainly the {\it threshold} stretching for base pair opening cannot be set \emph{a priori} merely according to the potential parameters, it should be rather estimated statistically through the mesoscopic model as discussed in Section 4.

Experiments show that bubbles of only a few base pairs are formed and undergo large amplitude thermal fluctuations which are highly localized \cite{bonnet1}. Moreover the lifetimes of open and closed states \cite{metz03} are sensitive to the solvent as separated strands can recombine at a rate which depends on the finite proton concentration in solution. This recombination event cannot be foreseen by the PB Hamiltonian which essentially deals with a DNA chain in a infinitely diluted solution.  Eq.~(\ref{eq:1}) does not account for re-closing
as the open strands can go infinitely apart with no energy cost due to the plateau in $V_M(y_n)$. Although the model may be improved by changing the shape of the potential for base pairs hydrogen bonds \cite{pey9}, a restriction of the configuration space is always required to keep the transverse stretchings within a finite range \cite{zhang}.
This is done in the path integral method as outlined in the next Section.

\section*{3. Path Integral Method}

In the finite temperature path integral formalism, the quantum statistical partition function $Z_Q$ is viewed as an analytical continuation of the quantum mechanical partition function to the imaginary time and it is calculated by integrating  a Boltzmann-like probability distribution over the particle path phase space

\begin{eqnarray}
Z_Q=\,\oint \mathfrak{D}x\exp\bigl[- A_Q\{x\}/\hbar\bigr]\,,\,
\label{eq:2}
\end{eqnarray}

where $A_Q\{x\}$ is the Euclidean action for the system, $\hbar$ is the Planck constant over $2\pi$, $\mathfrak{D}x$ is the integration measure and $\oint$ indicates that the paths $x(\tau)$ are closed trajectories \cite{io3}. $\tau$ is the imaginary time whose domain is $\tau \in [0, \beta]$,  $\beta$ being the inverse temperature \cite{mahan}.

In Ref.\cite{io}, I have proposed a path integral approach to the Hamiltonian in Eq.~(\ref{eq:1}) introducing the mapping of the real space interactions onto the imaginary time scale.
Accordingly, the transverse stretching $y_n$ is represented by a one dimensional path $x(\tau_i)$  obeying the periodicity property, $x(\tau_i)=\,x(\tau_i + \beta)$. In fact there are  $N + 1$ base pairs in Eq.~(\ref{eq:1}). Thus the period $\beta$ can be partitioned in $N_\tau + 1$ points $\tau_i$, each of them corresponding to a specific base pair $n$ along the DNA strand:

\begin{eqnarray}
& &y_n \rightarrow x(\tau_i), \, \, \nonumber
\\
& &n =\, 1\,...\,N  \,; \, \, i =\,1\,...\,N_\tau + 1 \, ,\,  \, \nonumber
\\
& &\tau_1 \equiv 0 \,; \, \,\tau_{N_\tau + 1} \equiv \beta \,.
\label{eq:3}
\end{eqnarray}

Thus, at any given temperature, the finite size system of $N + 1$ base pairs, is described by $N_\tau + 1$  paths taken at a specific $\tau_i$ along the time axis.
The presence of an extra base pair $y_0$ in Eq.~(\ref{eq:1}) is remedied by taking periodic boundary conditions, $y_0 = \, y_N$, which close the finite chain into a loop. Such conditions are easily incorporated in the path integral formalism as the path is a closed trajectory, $x(0)=\,x(\beta)$. Hence a molecule configuration is given by $N_\tau$ paths and, in the discrete imaginary time lattice, the separation between
nearest neighbors base pairs is $\Delta \tau =\,\beta / N_\tau$. Then, the mapping of Eq.~(\ref{eq:1}) on the time axis also requires:

\begin{eqnarray}
& &y_{n-1} \rightarrow x(\tau')\, \nonumber
\\
& &\tau' = \tau_i - \Delta \tau \, .\,
\label{eq:4}
\end{eqnarray}

Finally note that the real time derivative $\dot{y}_{n}$ maps onto the imaginary time derivative $\dot{x}(\tau)$ according to:

\begin{eqnarray}
{{d {y}_{n}} \over {dt}} \rightarrow (\nu \beta){{d x} \over {d\tau}}\,. \,
\label{eq:5}
\end{eqnarray}

This amounts to replace

\begin{eqnarray}
\hbar \rightarrow (\nu \beta)^{-1} \,,\,
\label{eq:6}
\end{eqnarray}

which is justified in the classical regime appropriate to DNA denaturation. The same replacement is done in order to solve the pseudo-Schr\"{o}dinger equation for a Morse potential \cite{landau} which is obtained from Eq.~(\ref{eq:1}) in the large $K$ limit (and $\rho=\,0$) \cite{pey1}.

Then, applying Eqs.~(\ref{eq:3}) - (\ref{eq:5}) to Eq.~(\ref{eq:1}), the classical partition function for the DNA molecule in one dimension reads

\begin{eqnarray}
& &Z_C=\,\oint \mathfrak{D}x\exp\bigl[- \beta A_C\{x\}\bigr]\, \nonumber
\\
& &A_C\{x\}=\, \sum_{i=\,1}^{N_\tau} \Bigl[{\mu \over 2}\dot{x}(\tau_i)^2 + V_S(x(\tau_i),x(\tau')) + V_M(x(\tau_i)) \Bigr] \, \nonumber
\\
& &x(\tau_i)=\, x_0 + \sum_{m=1}^{M_F}\Bigl[a_m \cos(\omega_m \tau_i) + b_m \sin(\omega_m \tau_i) \Bigr] \, \nonumber
\\ \,
& &\omega_m =\, {{2 m \pi} / {\beta}}\, \nonumber
\\
& &\oint \mathfrak{D}x\equiv {1 \over {\sqrt{2}\lambda_\mu}}\int_{-\Lambda^0_T}^{\Lambda^0_T} dx_0 \prod_{m=1}^{M_F}\Bigl({{m \pi} \over {\lambda_\mu}}\Bigr)^2 \times \, \nonumber
\\
& & \hskip 2cm \int_{-\Lambda_T}^{\Lambda_T} da_m \int_{-\Lambda_T}^{\Lambda_T} db_m \, \, , \, \nonumber
\\
\label{eq:4b}
\end{eqnarray}

where $A_C\{x\}$ is the classical Euclidean action. I take $N_\tau =\,100$ as the focus is here on short DNA fragments \cite{zocchi}.
The base pair path $x(\tau_i)$ can be Fourier expanded by virtue of the periodic boundary conditions, $M_F$ is  determined numerically and the measure of integration $\mathfrak{D}x$ is expressed through the coefficients $\{x_0 , a_m , b_m\}$. $\lambda_\mu$ is the thermal wavelength whose form in general depends on the model whether quantum \cite{io2b} or classical.  Due to Eq.~(\ref{eq:6}), ${\lambda_\mu}=\,\sqrt{{\pi } / {\beta K}}$.

The physical features of the path integral approach with the associated mapping technique can be synthesized as follows:

\emph{a)} One point $(x_0 , a_m , b_m)$ in the space of the Fourier coefficients selects an ensemble $\{x(\tau_i), i=\,1,\, N_\tau \}$ representing a molecule state with its peculiar base pair displacements. The state temperature dependence is monitored by varying $\beta$ as each $\tau_i \in [0,\beta]$.

\emph{b)} Spanning the Fourier coefficients phase space, at a fixed $\beta$, one builds a set of possible molecule configurations characterized by different choices of base pair stretchings for the same temperature. The higher is the number of configurations included in the path integral the more reliable is the computation of the DNA thermodynamics.

\emph{c)} The truncation of the configuration space mentioned in the Introduction occurs in the path integral method through the temperature dependent cutoffs ${\Lambda^0_T}$ and ${\Lambda_T}$ in the latter of Eq.~(\ref{eq:4b}) \cite{io2,io2a}. Both ${\Lambda^0_T}$  and ${\Lambda_T}$ are derived consistently with the requirement that the measure of integration normalizes the free particle action and with the physics contained in the model potential. This is explained in Subsection 4.3 where the possible cutoff model dependence of our thermodynamical results is also examined.
Paths $x(\tau_i) \sim 0$ represent the equilibrium configuration for the double helix corresponding to the minimum $V_M(x(\tau_i))$. Larger paths around the minimum should be incorporated in the computation however discarding \emph{ negative path amplitudes, smaller than $x_{min} \simeq -0.2{\AA}$} at high $T$, which are forbidden by the electrostatic repulsion between the sugar-phosphate backbones. The lower bound confinement for the $\{x(\tau_i)\}$, physically due to the hard core potential, also ensures that the base pair paths are self-avoiding at complementary sites along the strands. In fact base pair mates do not overlap. Then self-avoidance effects are included in the one dimensional model.

Also \emph{positive path amplitudes, larger than $x_{max} \simeq 6{\AA}$} at high $T$, should be discarded posing a threshold on the two strands separation. Inclusion of very large positive paths does not contribute to the free energy derivatives. Instead, an upper bound on the path displacements is consistent with strand recombination which may occur in the presence of a solvent.
Thus the range for the base pair stretchings, $x(\tau_i) \in [x_{min}, x_{max}]$, is qualitatively set according to the shape of $V_M(x(\tau_i))$.  Inside such range, the selection of the paths which indeed contribute to $Z_C$ is performed by means of a macroscopic constraint, the second law of thermodynamics. The strategy is the following:

Eq.~(\ref{eq:4b}) is computed, at an initial temperature $T_I$, for a given path ensemble defined (for any base pair) by the number of integration points over the Fourier coefficients, $N_{eff}$. The latter is temperature dependent. Then, at any larger $T$, the numerical code re-determines the contribution to $Z_C$ and calculates the derivatives of the free energy $F=\, -{\beta^{-1}}\ln Z_C$.

If, for a given $N_{eff}$, the growing entropy constraint is not fulfilled then the size of the path ensemble is increased. The procedure is reiterated until a {\it minimum number of paths} is found such that the entropy grows versus $T$.  This method sets the size of the ensemble whose paths satisfy boundary conditions and macroscopic physical constraints.   $N_{eff}$ is the $T$-dependent number of different trajectories followed by a single base pair stretching in the configuration space. As the procedure holds for any $\tau_i$, the total number of paths contributing to the thermodynamics is $N_\tau \times N_{eff}$. This value sets the overall system size. Numerical convergence has been found taking $N_{eff} \sim 1.2\times 10^5$ at $T_I=\,260K$  while there are no significant effects by further increasing the initial size of the path ensemble.

\section*{4. Results}

\subsection*{4.1. Fractions of Open Base Pairs}

The path integral method can characterize the melting through the computation, via Eq.~(\ref{eq:4b}), of the equilibrium thermodynamic properties. The latter provide a signature of the disruption of the base pair bonds.
In fact, the order parameter in homogeneous DNA denaturation is usually taken as the fraction of bound base
pairs ($f_b$) which is proportional to the internal energy of the chain, hence $- d f_b /d T$ is proportional to the specific heat.
Also in heterogeneous sequences the specific heat is an indicator of the melting as it displays sharp peaks at the temperatures where various fragments of the sequence open \cite{bresl}. While the definition of order parameter is less clear in this case \cite{joy08}, such peaks should be revealed by the behavior of $f_b$ . Theoretical models should therefore predict both macroscopic and microscopic denaturation features in the same temperature range although the overlap is not expected to be precise for heterogeneous DNA.

The melting transition is usually monitored by an increase in the UV absorption signal around $2600 \,{\AA}$. This is due to the fact that, when the double helix open and the bases move out of the stack, the corresponding electronic transitions are less screened. However spectroscopic measurements yield average fractions of open base pairs where the averaging is meant over an ensemble of molecules. This poses some questions to theorists and experimentalists:

\emph{i)} There is some intrinsic arbitrariness in the definition of an \emph{open} base pair as one needs to assume a threshold for the elongations beyond which the pair dissociates.

\emph{ii)} When the UV signal measures that half of the base pairs are open, this may indicate \emph{either} that half of the molecules are open and half are closed \emph{or} that all molecules are half-open. Accordingly these methods cannot distinguish intermediate states for a single molecule configuration which, instead, would be quite interesting in order to understand the nature of the melting transition. New techniques based on quenching of single strands are becoming available to trap intermediate states \cite{zocchi2}.

Eq.~(\ref{eq:4b}) accounts for an ensemble of different configurations for one molecule. Statistical averages are therefore carried out over such ensemble and the mean elongation for the $i-th$ base pair reads:

\begin{eqnarray}
& &< x(\tau_i) >=\,Z_C^{-1}\oint \mathfrak{D}x x(\tau_i) \exp\bigl[- \beta A_C\{x\}\bigr] \,, \,
\label{eq:7}
\end{eqnarray}

where the integral includes those {\it good paths} which are selected by the method described in the previous Section.
Introducing a {\it threshold} $\zeta$ , a base pair is open if: $<x(\tau_i)>\, \geq\, \zeta$. Thus we are able to compute the profiles $<x(\tau_i)>\, - \zeta$ as a function of the site index $i$ for different temperatures \cite{krueg}.
$\zeta$ is an arbitrary parameter generally depending both on the length and on the sequence of the fragment: for the present model it may be reasonably taken \cite{pey3,zhang} in the range $\sim [0.5 - 2]{\AA}$  thus checking whether, for a given $\zeta$, there is a significant fraction of base pairs which dissociate close to some temperature. Such $T$ values may then be signatures of denaturation steps.
Too small $\zeta$ would lead to the wrong conclusion that all base pairs are already open at too low temperatures. Too large $\zeta$ would prevent us to follow the multistep melting of different regions of the chain at different temperatures.

As the UV signal changes quite abruptly when base pairs dissociate, Heaviside step function $\theta(\bullet)$ is generally used to define the fraction of open base pairs $f=\,1 - f_b$. The latter is expressed in terms of  Eq.~(\ref{eq:7}) as:

\begin{eqnarray}
& &f =\, {1 \over {N_\tau}}\sum_{i=1}^{N_\tau} \theta\bigl(< x(\tau_i) > - \zeta \bigr) \, . \,
\label{eq:8}
\end{eqnarray}

Eq.~(\ref{eq:8}) is consistent with the definition adopted in Monte Carlo simulations \cite{ares} and dynamical mesoscopic models \cite{joy06}.

Alternatively, in the case of a very short experimental signal which is not related to $< x(\tau_i) >$,  one may rather perform the  statistical averages  $< \theta\bigl(x(\tau_i)  - \zeta \bigr) >$ and define the fraction of open base pairs $f'$ as

\begin{eqnarray}
& &f' =\, {1 \over {N_\tau}}\sum_{i=1}^{N_\tau} \bigl< \theta( x(\tau_i)  - \zeta ) \bigr >\, . \,
\label{eq:9}
\end{eqnarray}

While there is no a priori reason to prefer Eq.~(\ref{eq:8}) or Eq.~(\ref{eq:9}), I compute here both expressions to verify which one is more suitable to capture the multistep denaturation of an heterogeneous molecule in the framework of the path integral formalism.

I consider a $N_\tau =\,100$ fragment with an extended AT-substrate in the right side and a slight predominance of GC-pairs on the first (from left) 48 sites. The left part of the segment has the same sequence as the L48AS fragment of Ref.\cite{zocchi2} although our model cannot distinguish a configuration \emph{GC- followed by GC-} along the backbone from a \emph{GC- followed by CG-pair}. Experimentally these configurations present some differences in the free energy stackings \cite{yakov}. The sequence is:

\begin{eqnarray}
& &GC + 6AT +  GC + 13AT + 8GC + AT + 4GC + \nonumber
\\
& &AT + 4GC + AT + 8GC + [49-100]AT \, , \,\nonumber
\\
\label{eq:10}
\end{eqnarray}

for which bubbles may open  in the leftmost and, more likely, in the right part of the chain.

\begin{figure}
\includegraphics[height=7.0cm,angle=-90]{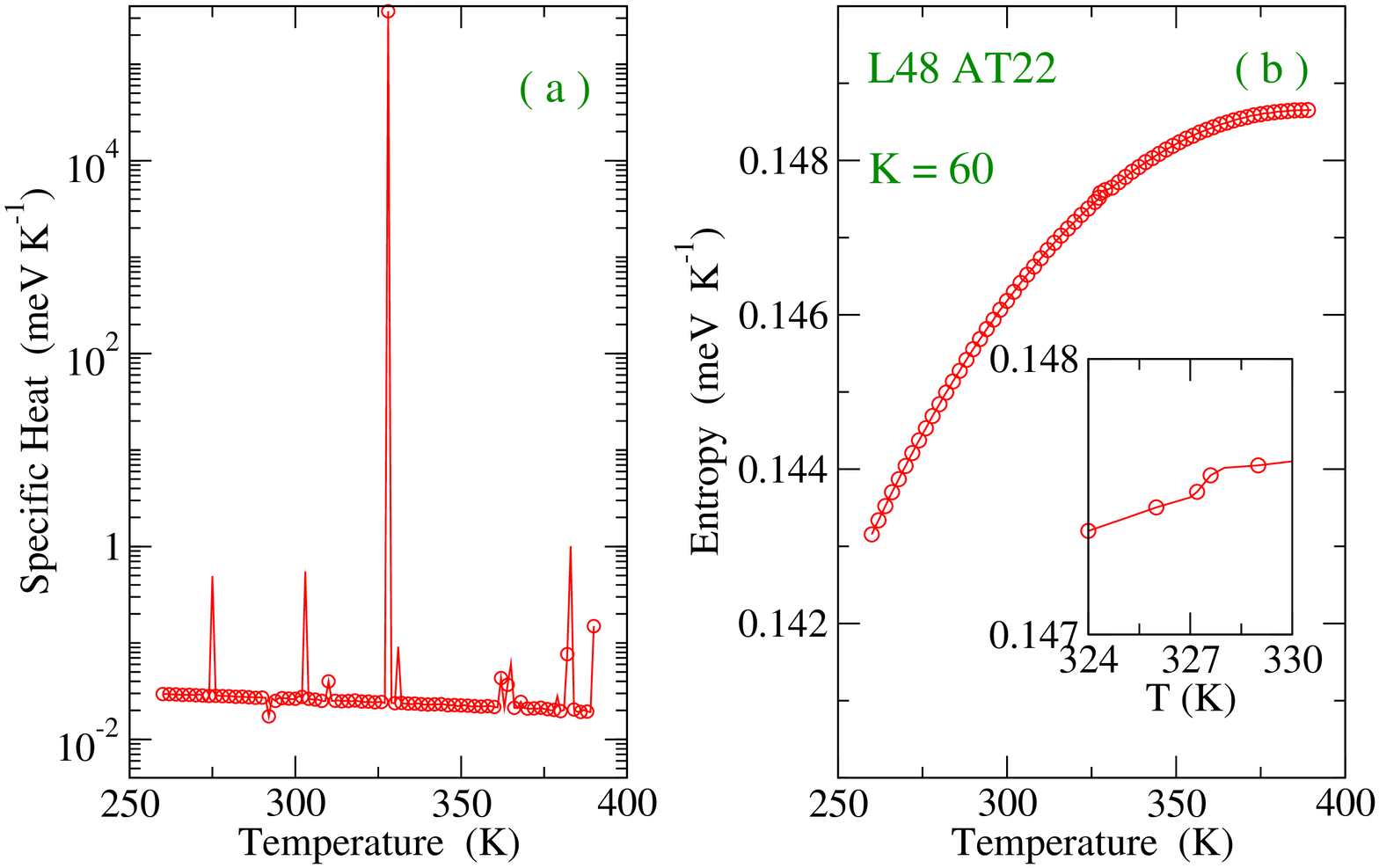}
\includegraphics[height=7.0cm,angle=-90]{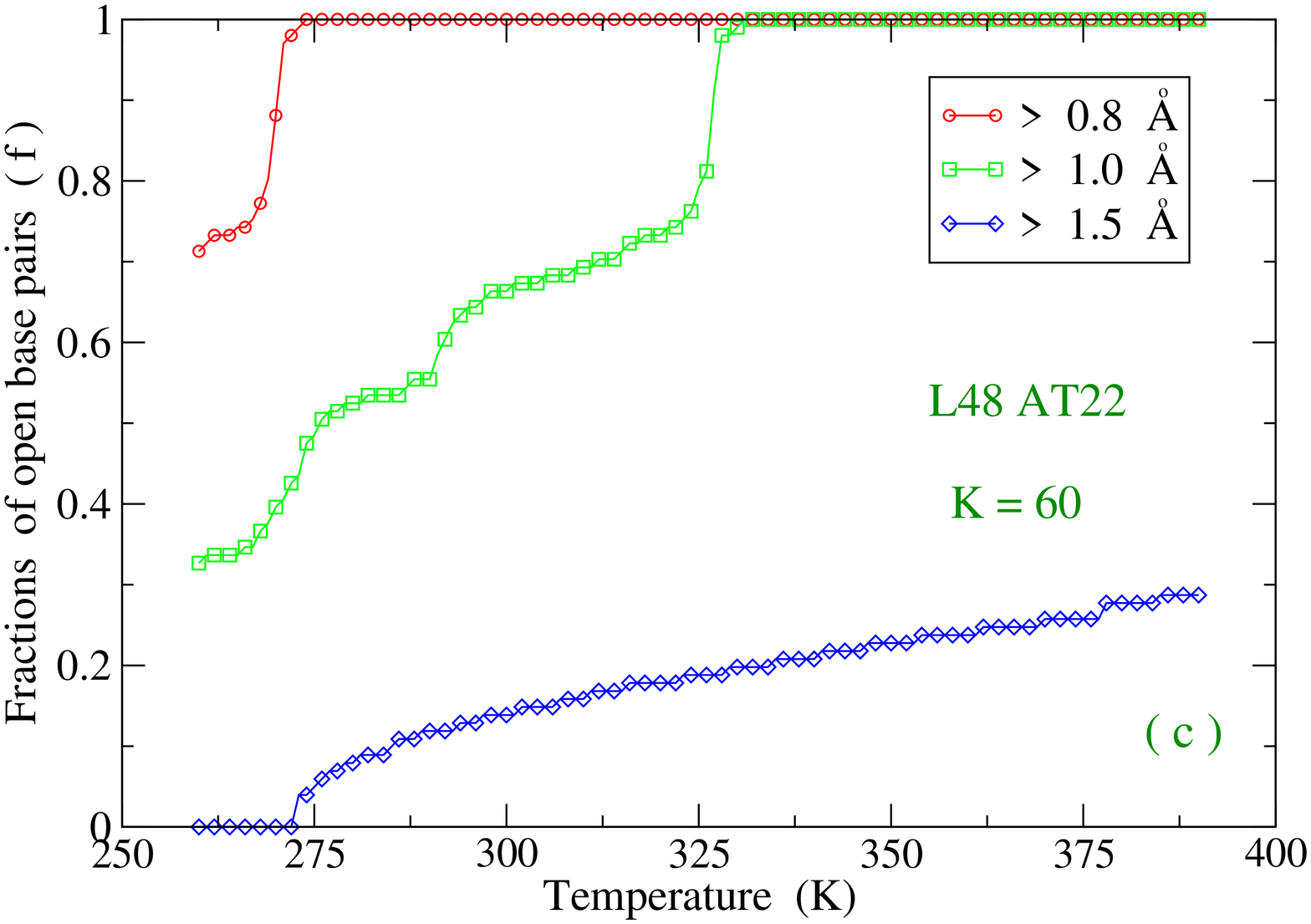}
\includegraphics[height=7.0cm,angle=-90]{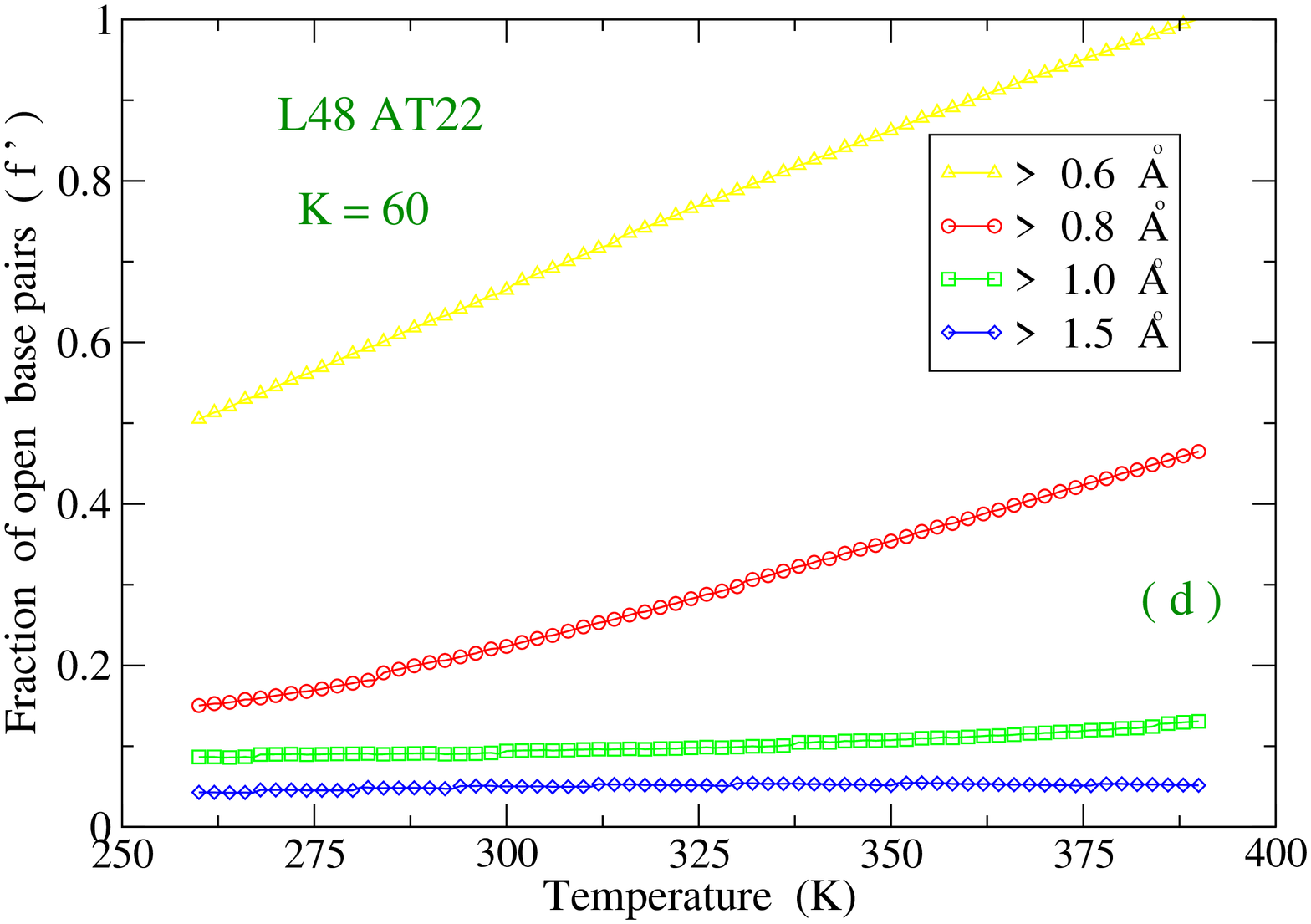}
\caption{\label{fig:1}(Color online) (a) Specific Heat and (b) Entropy for the fragment in Eq.~(\ref{eq:10}) and stacking coupling $K=\,60 \, meV {\AA}^{-2}$.  (c) Fraction of open base pairs according to Eq.~(\ref{eq:8}). (d) Fraction of open base pairs according to Eq.~(\ref{eq:9}).}
\end{figure}

The thermodynamics results are presented in Fig.~\ref{fig:1} for a backbone stiffness $K=\,60 \, meV {\AA}^{-2}$ which corresponds to a weak intra-strand coupling. The entropy (Fig.1(b)) is a continuous function of temperature with a slight kink at around $T =\,327K$ (magnified in the inset) which is mirrored by a pronounced peak in the specific heat plot (Fig.1(a)). The specific heat also displays an array of minor side peaks at $T \sim 275, \, 300, \, 380K$. The size of the entropy step is $\sim 10^{-4} meV K^{-1}$, much smaller than the value $0.0646 meV K^{-1}$ found in Ref.\cite{theo}, albeit for homogeneous DNA. Even bigger melting entropies have been predicted for long chains of heterogeneous DNA using extended transfer matrix approach for the PB model but with a very large non linear $\rho$ \cite{theo1}. Instead, the smooth entropy behavior found in the path integral approach is more in line with the transfer matrix analysis for a sequence of 100 base pairs of the Joyeux-Buyukdagli model \cite{joy05,joy07} which also estimates reduced melting entropy (with respect to the PB model) at the thermodynamic limit.

Look now at Fig.1(c) which reports $f$ in Eq.~(\ref{eq:8}) for three selected values of $\zeta$: at $T \sim 275K$, roughly $8\%$ of the average base pair stretchings become larger than $\zeta=\, 1.5 \,{\AA}$, $50\%$ become larger than $\zeta=\, 1 \,{\AA}$ and all average paths become larger than $\zeta=\, 0.8 \,{\AA}$. At $T \sim 295K$, there is a further enhancement ($15 \%$ more) in the fraction of paths exceeding $\zeta=\, 1\, {\AA}$ while $f$ grows smoothly for $\zeta=\, 1.5 \, {\AA}$. At $T \sim 327K$, all average paths become larger than $\zeta=\, 1 \,{\AA}$ consistently with the main peak in the specific heat.
At $T \sim 380K$, $f$ with $\zeta=\, 1.5 {\AA}$ has another small increase. The fact that the $f-$ plots display step-like enhancements at various $T$ also explains why the specific heat peaks are very narrow in temperature.

Using $f'$ in Eq.~(\ref{eq:9}), we would not be able to capture an evident correspondence between base pair dissociations and specific heat peaks: Fig.1(d) makes clear in fact that $f'$ grows steadily for any $\zeta$. Thus, it seems that the averaging method in $f$ is more suitable to account for a multistep melting at least for the one-molecule system here treated. Our conclusion is in qualitative agreement with Refs.\cite{joy09,ares} although the investigated sequences differ. Instead, in Ref.\cite{pey3}, the expression in Eq.~(\ref{eq:9}) has been preferred.  Note that, for a given $\zeta$, $f' \ll f$ at any $T$.

The overall picture emerging from Fig.~\ref{fig:1} is that of a transition driven by the AT-rich portions of the fragment. Such transition is a smooth crossover (as shown by the entropy plot) mainly occurring in the range $T \sim [275 \,- \,327]K$ within which all pair displacements become larger than $\zeta=\, 1 \,{\AA}$. Certainly, this does not imply that the molecule is open at $T \sim 327K$ as many more paths, mainly associated to the GC-pairs, broaden at higher $T$ providing a further entropic gain. In this regard, a recent neutron scattering study has pointed out how significant clusters of base pairs continue to be closed inside the denaturation regime \cite{theo2}.
In fact the entropy still grows above the main denaturation peak although the rate of the growth gradually decreases. While we cannot define a threshold for the complete strands separation within the considered temperature range, we observe that $70\%$ of the path stretchings belong to the range $[1 \, - \, 1.5]{\AA}$ at the upper value $T = 390K$  while  $30\%$ are  larger.

\begin{figure}
\includegraphics[height=7.0cm,angle=-90]{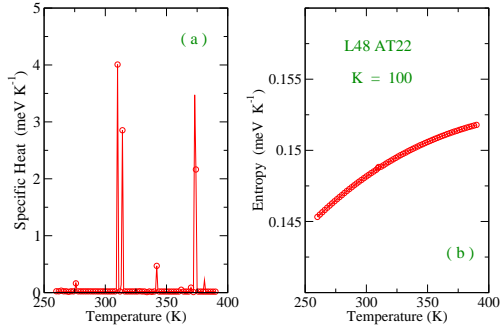}
\includegraphics[height=7.0cm,angle=-90]{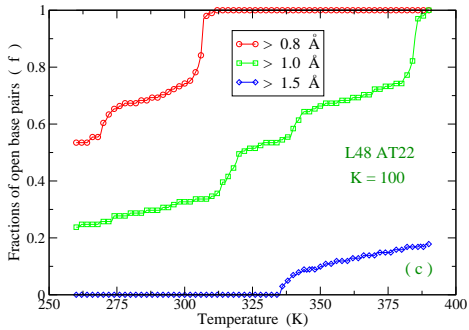}
\includegraphics[height=7.0cm,angle=-90]{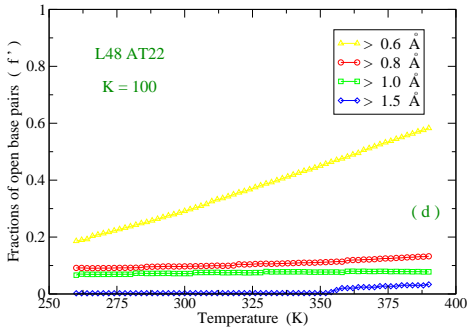}
\caption{\label{fig:2}(Color online) (a) Specific Heat and (b) Entropy for the fragment in Eq.~(\ref{eq:10}) and stacking coupling $K=\,100 \, meV {\AA}^{-2}$.  (c) Fraction of open base pairs according to Eq.~(\ref{eq:8}). (d) Fraction of open base pairs according to Eq.~(\ref{eq:9}).}
\end{figure}

The calculations for a stacking coupling $K=\,100 \, meV {\AA}^{-2}$, are presented in Fig.~\ref{fig:2}. The entropy plot (Fig.2(b)) is even smoother than in the previous case so that the specific heat peaks (Fig.2(a)) are much less sharp.
Hardening the backbone stiffness has the main effect to shift the melting signatures towards higher temperatures. Thus, see Fig.2(c), all average base pairs become larger than $\zeta=\, 0.8 \,{\AA}$ at $T \sim 310K$, a $35K$ increase with respect to Fig.1(c). At about the same $T$ there is also a $20\%$ increment in the fraction of average paths larger than $\zeta=\, 1 \,{\AA}$. For the same value, $f$ shows two more step-like features at $T \sim 340K$ and $T \sim 385K$.  Again we find an overall correspondence with the specific heat peaks locations at $T \sim 310K$, $T \sim 340K$ and $T \sim 375K$ respectively.
Also in this case $f'$ grows smoothly, see Fig.1(d), providing no indication of bubble formation along the strands.

\subsection*{4.2. Cooperativity}

\begin{figure}
\includegraphics[height=7.0cm,angle=-90]{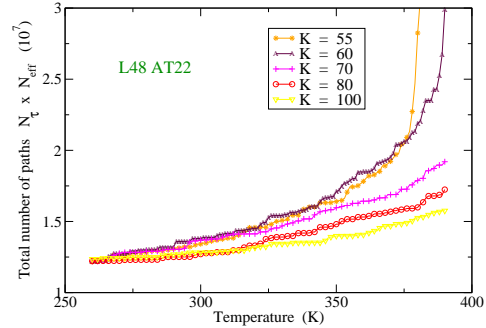}
\caption{\label{fig:3}(Color online) Total number of paths used in the computation of the thermodynamics of the DNA sequence for five values of the stacking parameter $K$ in units $meV {\AA}^{-2}$. }
\end{figure}

Theories for DNA have focussed since long on the cooperative character of the melting transition \cite{poland,proh}. The interplay between stacking coupling and cooperativity in the model of Eq.~(\ref{eq:1}) has been pointed out in Section 2. In the path integral method the degree of cooperativity is measured by the parameter $N_\tau \times N_{eff}$ which yields the total number of trajectories contributing to Eq.~(\ref{eq:4b}) at a given $T$. This is set by the numerical code as described in Section 3. I emphasize that $N_\tau \times N_{eff}$ represents (for any $T$) the \emph{minimum} number of paths for which the second law of thermodynamics is fulfilled. Once such minimum value is determined there is no need to further increase the size of the paths ensemble. In Fig.~\ref{fig:3}, $N_\tau \times N_{eff}$ is plotted versus $T$  for the sequence in Eq.~(\ref{eq:10}) as a function of the stiffness $K$.

While the computation starts with $1.2 \times 10^7$ paths for any $K$ at the lower bound of the $T$ range, the renormalization of
$N_\tau \times N_{eff}$ along the $T$-axis remarkably depends on $K$. At $T \sim 390K$, about $3 \times 10^7$ paths are included in $Z_C$ for the case $K=\,60 \, meV {\AA}^{-2}$ whereas such number drops by a factor two in the case $K=\,100 \, meV {\AA}^{-2}$. For the former case, also note the step-like enhancement at $T \sim 327K$ consistent with the main peak in Fig. 1(a). An enhanced stiffness reduces the flexibility of the strand and inhibits the bubble formation thus shifting the occurrence of the melting steps at higher $T$. Here we see how the intra-strand stacking is key to determine the inter-strands opening probabilities. The lowest $K$ values here assumed are in the range of those usually taken in nonlinear PB Hamiltonian models \cite{theo}.

\subsection*{4.3. Path Amplitudes Cutoffs }

The cutoffs ${\Lambda^0_T}$ and ${\Lambda_T}$ in Eq.~(\ref{eq:4b}) permit to build an ensemble of paths whose amplitude is temperature dependent. To derive analytically the form of the $T-$ dependence I observe that the integration measure normalizes the free particle action:

\begin{eqnarray}
\oint \mathfrak{D}x(\tau)\exp\Bigl[- \int_0^\beta d\tau {\mu \over 2}\dot{x}(\tau)^2  \Bigr] =\,1.
\label{eq:11} \,
\end{eqnarray}

By inserting the Fourier path expansion in Eq.~(\ref{eq:4b}), the l.h.s. of Eq.~(\ref{eq:11}) transforms into a product of Gau{\ss}ian integrals which yield the mathematical criteria to set ${\Lambda^0_T}$ and ${\Lambda_T}$ through the conditions:

\begin{eqnarray}
& &{\sqrt{2} \over {\lambda_\mu}}\int_{0}^{\Lambda^0_T} dx_0 =\, 1 \, \nonumber
\\
& &{2 \over {\sqrt{\pi}}} \int_{0}^U ds_m \exp(-s_m^2) =\, 1 \, \nonumber
\\
& &s_m^2 \equiv \,{{m^2 \pi^3 a_m^2} \over {\lambda^2_\mu}}. \, \nonumber
\\
\label{eq:12}
\end{eqnarray}

with $U$ being a dimensionless cutoff which can be set numerically by taking, for instance, a series expansion for the Gau{\ss}ian integral \cite{grad}.
From Eqs.~(\ref{eq:12}), it follows that:

\begin{eqnarray}
& &{\Lambda^0_T}=\,\lambda_\mu/\sqrt{2} \, \nonumber
\\
& &\Lambda_T =\,{{U \lambda_\mu}  \over {m \pi^{3/2}}},
\,
\label{eq:13}
\end{eqnarray}

hence, the path amplitude cutoffs display a $\propto \sqrt{T}$ behavior.

\begin{figure}
\includegraphics[height=7.0cm,angle=-90]{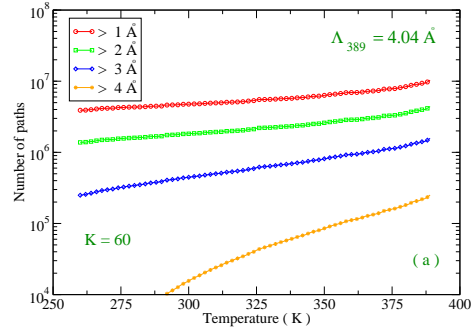}
\includegraphics[height=7.0cm,angle=-90]{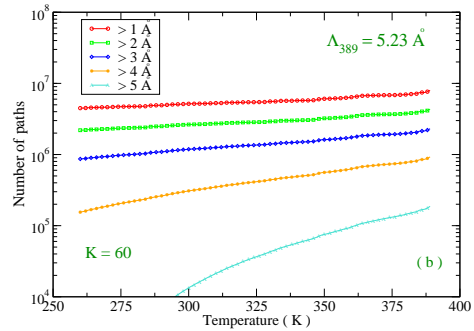}
\includegraphics[height=7.0cm,angle=-90]{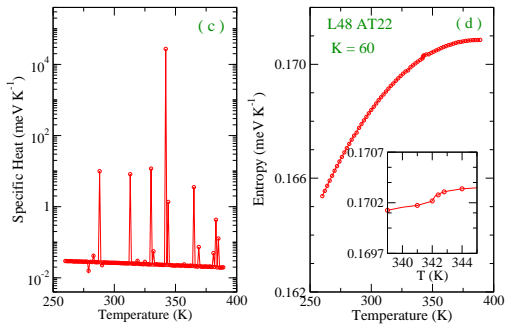}
\caption{\label{fig:4}(Color online) (a) Number of paths larger than $1,2,3,4 \,{\AA}$ taking in Eq.~(\ref{eq:4b}) the same cutoff, $4.04\,{\AA}$ at $T=\,389K$,  as in Fig.~\ref{fig:1} and ~\ref{fig:2}. (b) Number of paths larger than $1,2,3,4,5 \,{\AA}$ assuming a larger cutoff, $5.23\,{\AA}$ at $T=\,389K$, in Eq.~(\ref{eq:4b}).
(c) Specific Heat and (d) Entropy calculated with the $\Lambda_T$ value given in (b). $K$ in units $meV {\AA}^{-2}$. }
\end{figure}

There is however some freedom in the choice of $U$, that is one may tune the cutoff within a range which essentially fulfills both Eq.~(\ref{eq:12}) and the physical requirements of the problem in Eq.~(\ref{eq:4b}) emphasized in the previous Section. In this way we have a mathematically consistent approach to explore the hydrogen bonds potential plateau and include a set of base pair paths suitable to the specific temperature of the system.
All the results presented so far have been obtained taking $U=\,17$ that implies $\Lambda_{389K}=\,4.04 {\AA}$ for the first Fourier coefficient. Accordingly the path displacements are not larger than $\sim 5{\AA}$ inside the investigated temperature range. Fig.~\ref{fig:4}(a) shows the total number of paths larger than $1,2,3,4 \,{\AA}$ respectively, used in the computation of Fig.~\ref{fig:1}. Note that a significant number of paths, $\sim 10^5$, larger than $4 \,{\AA}$ \emph{but smaller than} $5 \,{\AA}$ contributes to $Z_C$ at $T \sim 350K$.

How do the thermodynamical properties depend on the choice of $U$? Taking $U=\,22$, which means $\Lambda_{389K}=\,5.23 {\AA}$, I incorporate even broader base pair displacements as made evident by Fig.~\ref{fig:4}(b): at $T \sim 350K$ there are now about $10^5$ paths, larger than $5 \,{\AA}$ \emph{but smaller than} $6 \,{\AA}$. Specific heat and entropy are plotted in Fig.~\ref{fig:4}(c) and Fig.~\ref{fig:4}(d) respectively. A remodulation in the specific heat peaks occurs with respect to Fig.~\ref{fig:1}(a): three sizeable side peaks (rather than two) show up in the specific heat at temperatures lower than the main peak which is now located at $T=\,342K$. Some portions of the fragment tend to open before the main transition temperature is attained while the latter is shifted slightly upwards in comparison with the case of Fig.~\ref{fig:1}(a). As a main result, the entropy jump at $T=\,342K$ is again small and of order $10^{-4} meV K^{-1}$, see inset in Fig.~\ref{fig:4}(d). Likewise taking $U=\,100$, $\Lambda_{389K}=\,23.79 {\AA}$, thus confirming that the smooth character of the denaturation transition is not an artifact of the truncation procedure in the path configuration space. In the path integral method, the cutoff dependence appears altogether not so crucial as the ensemble size $N_\tau \times N_{eff}$ is large even for short sequences.

\section*{5. Conclusion}

I have applied the path integral formalism to an heterogeneous DNA sequence modeled through the nonlinear Peyrard-Bishop Hamiltonian which captures the main interactions of the double strands configuration. The method allows us to compute the thermodynamic properties incorporating all base pairs thermal fluctuations that are particularly relevant in a short fragment. The energetic gain associated to the (bounded) double strands molecule competes with the entropic gain due to the large number of configurations available once the two strands begin to dissociate. The computation refers to a temperature window in which denaturation is expected to occur and includes an high number of possible trajectories for the base pair stretchings. Such number is set at any temperature on the base of physical criteria included in the numerical code. First,  the path configuration space is truncated consistently with the requirements of the hydrogen bonds model potential. Second, the allowed base pair paths are selected by simply requiring that the entropy should have a positive temperature derivative. Such requirement does not introduce any constraint on the shape of the entropy itself which always appears as a continuous function of $T$. Thus the denaturation manifests as a smooth crossover taking place in multisteps as the average stretchings of the \emph{adenine-thymine} base pairs are larger than those of the \emph{guanine-cytosine} base pairs.

The work has focused on the quantitative effects of the backbone stiffness and, in particular, it has been shown that such parameters determines the size of the path ensemble included in the computation of the partition function. This points to a relevant role of the stacking coupling on the degree of cooperativity of the molecule. A weak stacking induces flexibility of the double strand structure and allows an high number of paths to participate to the multistep melting. The importance of anharmonicity in the intra-strand potential with regard to the melting transition is controversial and has been widely discussed in the literature: we recognize that the anharmonic stacking drives a cooperative behavior of the base pairs in the PB Hamiltonian model but the effects of the coupling on the denaturation features are due to the harmonic parameter $K$ rather than to the anharmonic force constants. It is the value of $K$ which sets the location of the melting steps along the $T-$ axis and determines the fractions of open base pairs. However varying the stacking coupling does not change the character of the transition which remains smooth as in homogeneous fragments. In particular, I find that the entropy jump associated to the main melting transition is very small due to fluctuational effects which are strong in a short chain, whereas the fractions of open base pairs display step-like enhancements in remarkable correspondence with the peaks of the specific heat. This investigation also sheds light on the averaging procedure to estimate such fractions for a molecule existing in many different configurations.

Among the main advantages of the path integral method there is certainly the capability to include a large-size ensemble of base pair fluctuations within a self-contained computational time.
While this study refers to a one-dimensional system and cannot account for the helicoidal structure of the molecule, the method seems promising also for extensions to the three dimensional space. This would also permit to fully incorporate excluded volume effects which have been here only partially considered.

\end{document}